\documentclass{ws-ijmpcs}
\usepackage{graphicx}
\usepackage{float}

\def\trimmarks{}

\begin{document}

\markboth{Alexei Prokudin}
{QCD Evolution Workshop: Introduction}

%
\catchline{}{}{}{}{}
%

\title{QCD Evolution Workshop: Introduction 
}


\author{Alexei Prokudin}

\address{Thomas Jefferson National Accelerator Facility,\\
   Newport News, VA 23606, U.S.A. \\
prokudin@jlab.org}

\maketitle

\begin{history}
\end{history}

\begin{abstract}
The introduction talk given at the beginning of QCD Evolution workshop
held in Thomas Jefferson National
Accelerator Facility (Jefferson Lab) on May 14 -17, 2012. 
\end{abstract}

\ccode{PACS numbers: 11.10.-z,12.38.-t, 12.38.Bx}

\section*{Past, present, and future}

QCD Evolution workshop is an annual workshop traditionally held in Jefferson Lab
and aimed at bringing together theoretical and experimental physicists working
in the domain of nuclear and high energy physics and employing QCD in their research.

This year we\footnote{Organizing committee: \\
Alexei Prokudin (JLab) Chair \\
Anatoly Radyushkin (ODU\&JLab) \\
Ian Balitsky (ODU\&JLab) \\
Leonard Gamberg (PSU)\\
Harut Avakian (JLab)} organize the second edition\footnote{http://www.jlab.org/conferences/qcd2012/program.html}  of the workshop 
following the success of the 
first one\footnote{http://www.jlab.org/conferences/QCDEvolution/program.html} organized in 2011.
The proceedings of the first workshop were published in Ref.~[\refcite{proceedings}].  

With the advent of quark parton model and Bjorken scaling in 1960s the theoretical and 
experimental studies of the hadron structure and in particular the proton's one became an important part of nuclear physics agenda
throughout the world.
Indeed by studying the proton we understand the underlying nature of Quantum Chromo Dynamics (QCD) -- the theory that
describes the proton as bound system of quarks and gluons. Quarks are confined inside the proton i.e.
thy are not observed as free asymptotic states in detectors, however asymptotic freedom of QCD, the fact that
at short distances quarks behave as free particles, allows
to study structure of the proton at small distances by varying, for example, virtuality $Q^2$ of the
incident photon in Deep Inelastic Scattering. 

Protons are used as discovery tool
in several facilities including Large Hadron Collider and precise knowledge of its structure becomes 
an essential ingredient of the discovery potential of such facilities. 

Jefferson Lab is accomplishing 12 GeV upgrade project [\refcite{Dudek:2012vr}] which is due to be in operation stage in 
2015 and will enable us to look with an 
unprecedented precision at the nucleon structure in the region where valence quarks are dominant in
 nucleon's waive function. Such precision is needed for better understanding of nature of the nucleon
as a many body relativistic system in terms of internal dynamics.

\begin{figure}[H]
\centering \includegraphics[width = 0.45\textwidth]{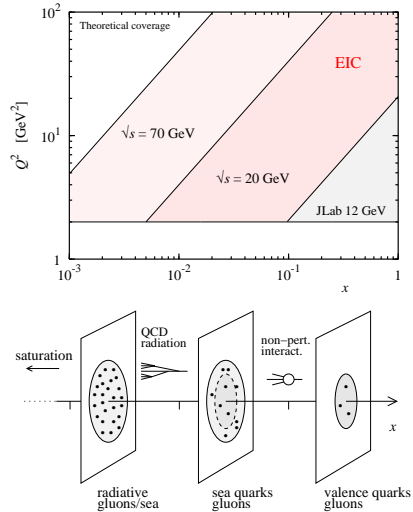}
\caption{(Top) Kinematic coverage in $x$ and $Q^2$ in ep scattering
experiments with JLab 12 GeV and a medium-energy EIC of
CM energy $\sqrt{s}$ = 20 and 70 GeV. The minimum momentum
transfer here was chosen as $Q^2_{min}$ = 2 GeV$^2$. (Bottom) Components 
of the nucleon wave function probed in scattering experiments
at different $x$.\label{fig1}}
\end{figure}

Looking forward in future one would like to study the dynamical origin of quarks ad gluons in the region where sea quarks
and gluons start dominating nucleon's waive function. This can be achieved by constructing a new facility -- polarized Electron Ion Collider 
[\refcite{Dudek:2012vr,Accardi:2011mz,Abeyratne:2012ah}]
 or EIC with variable center-of-mass energy $\sqrt{s}$ $\sim$ 20 --70 GeV and luminosity $\sim$ $10^{34}$ cm$^{-2}$ s$^{-1}$
 that  would be uniquely suited to address several outstanding
questions of Quantum Chromodynamics (QCD) and the microscopic structure of hadrons and nuclei. See Fig.~\ref{fig1}
\footnote{The plot is from Ref.~[\refcite{Accardi:2011mz}]. See Ref.~[\refcite{Accardi:2011mz}] for details on nuclear physics 
opportunities at a medium-energy EIC.} where kinematical range of JLab and EIC are
compared as functions of Bjorken-x and $Q^2$.

Spin and polarization measurements has been playing a crucial role in our understanding of nucleon's properties throughout many decades.
Since famous ``Spin crisis''[\refcite{Aubert:1985fx,Leader:1988vd}] of 1980's we learned that quark spins do not account for the
full spin of the nucleon. Given the later observation that the contribution of the gluon spin to that of the nucleon
could be rather small [\refcite{deFlorian:2011ia}] one concludes that a static picture of the nucleon with quarks 
in s-states does not account for the complexity of the parton dynamics. Orbital motion of quarks and gluons plays important role
in our understanding of the nucleon's structure. 

In recent years the description of the nucleon's spin and momentum 
structure given in terms of partonic sub-structure has led to rapid development of QCD theory. In hard semi-inclusive processes
 involving non-collinear dynamics these structures are described 
by Transverse Momentum Parton distributions and fragmentation functions (TMD-PDFs and TMD-FFs, or jointly TMDs). TMDs 
depend both on Bjorken--$x$ and transverse motion of partons $\bf k_T$ thus making them sensitive to Orbital Angular
Momentum of quarks and gluons.
The transverse degrees of freedom also play a crucial role in high energy collider experiments 
through so called Efremov-Teryaev-Qiu-Sterman matrix elements [\refcite{Efremov:1981sh,Efremov:1984ip,Qiu:1991pp}] i.e. multi-parton correlations. 

In more exclusive processes such as Deep Virtual Compton Scattering or Exclusive Vector Meson Electro production
one encounters so-called Generalized Parton Distributions (GPDs) that, by Fourier transform over
transferred momentum $t$, depend additionally to usual Bjorken--$x$ 
on position of partons in coordinate space.

\begin{figure}[H]
\centering \includegraphics[width = 0.53\textwidth]{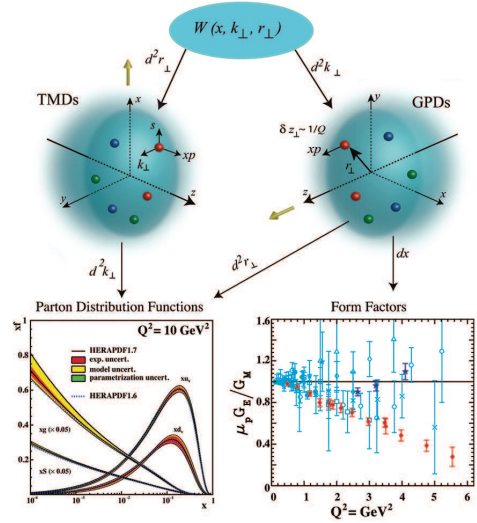}
\caption{Wigner distribution and relation to Generalized Parton Distributions and Transverse Momentum Dependent Distributions. Parton distributions 
and form factors can be related to GPDs and TMDs. \label{fig2}}
\end{figure}

Our grasp of the fundamental quark and gluon structure of the nucleon founded on QCD is still in its infancy, yet 
we lack a detailed understanding of these objects from first principles. Nevertheless, in the last ten years we have entered 
a new era where a framework suitable for a comprehensive and quantitative approach to the description of nucleon structure has emerged 
[\refcite{Ji:2003ak}]. 
In this framework our knowledge of nucleon structure is encoded in the Wigner distributions of the constituents, 
a quantum mechanical concept, introduced in 1932 [\refcite{Wigner:1932eb}]. 
From the Wigner distributions, see Fig.~\ref{fig2}\footnote{The plot is from Ref.~[\refcite{Dudek:2012vr}]}, a natural interpretation of measured observables is provided through the construction of 
its integrated ``slices'' or projections which are in fact Generalized Parton Distributions and Transverse Momentum Dependent distributions.

\section*{QCD Evolution}

The nucleon in QCD represents a dynamical system of
fascinating complexity. In the rest frame it may be viewed
as an ensemble of interacting color fields, coupled in an
intricate way to the vacuum fluctuations that govern the
effective dynamics at distances $\sim$ 1 fm.   A
complementary description emerges when one considers a
nucleon that moves fast, with a momentum much larger
than that of the typical vacuum fluctuations. In this limit
the nucleon’s color fields can be projected on elementary
quanta with point-particle characteristics (partons), and
the nucleon becomes a many-body system of quarks and
gluons. As such it can be described by a wave function, in
much the same way as many-body systems in nuclear or
condensed matter physics. In contrast to these
non-relativistic systems, in QCD the number of point-like
constituents is not fixed, as they constantly undergo creation/
annihilation processes mediated by QCD interactions,
reflecting the essentially relativistic nature of the
dynamics.

Accordingly the QCD evolution that governs content of the nucleon 
is interpreted differently in different frames. 

If one considers evolution of parton
densities with energy than the appropriate frame is so-called dipole frame
in which virtual photon fluctuates into a color dipole (quark--antiquark pair) and this dipole
interacts with target nucleon. Corresponding evolution is governed by Balitsky-Fadin-Kuraev-Lipatov (BFKL)
evolution equation [\refcite{Kuraev:1977fs,Balitsky:1978ic}]. The non linear regime of this evolution is described via Balitsky equation 
[\refcite{Balitsky:1995ub}]
Balitsky-Kovchegov equation in large $N_c$ limit 
(BK) [\refcite{Balitsky:1995ub,Kovchegov:1999yj,Kovchegov:1999ua}] and JIMWLK evolution equations 
[\refcite{JalilianMarian:1997gr,Iancu:2000hn,Ferreiro:2001qy}]. Subsequently the system will pass from dilute to dense 
regime of QCD and to predicted but yet to be observed regime of saturation of gluon densities. Geometrical
scaling of structure functions at low-$x$ observed at HERA [\refcite{Schildknecht:2000zt}] is an indication of this regime to
take place. Note that the resolution scale that is defined
by the virtuality of the photon $Q^2$ is fixed in this case.

DGLAP equation describes the evolution of densities as function of $Q^2$ at given energy scale or rapidity $y$. Infinite 
Momentum Frame (the frame in which the target nucleon moves with infinite momentum and thus along light-cone) is suitable for interpretation in this case.
Fluctuations of incident photon into quark-antiquark pairs are suppressed and the photon probes ``frozen'' partonic states inside
of the nucleon.
Gluon radiation in the available phase space produces multiple quark, antiquark and gluon states that are responsible for the
growth of parton densities in low-$x$ region. Not that the resolution scale $Q^2$ increases and thus the distance at which the
states are probes and the ``effective size'' of partons diminishes.

\begin{figure}[H]
\centering \includegraphics[width = 0.4\textwidth]{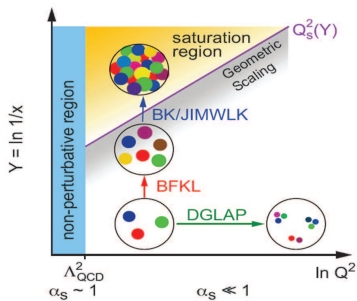}
\caption{Evolution of parton densities can be considered either in energy/rapidity $y$ (BFKL, BK and JIMWLK equations) or in virtuality of the photon $Q^2$ 
(DGLAP equation). The system will go from dilute towards dense regime and undergo transition to saturation region which is characterized by saturation scale $Q^2_s(y)$.
\label{fig3}}
\end{figure}

Evolution of parton densities consequently go from so-called dilute to dense regime and one expect that the phenomenon of saturation 
(recombination of gluons due to non abelian nature of QCD) will take place at some characteristic scale $Q^2_s(y)$. See Fig.~\ref{fig3} for
representation of different evolutions.

Evolution of Transverse Momentum Dependent distributions is an emerging subject of nuclear theory. The details of
TMD factorization were derived in Ref.~[\refcite{CollinsBook}] and successfully implemented in Refs.~[\refcite{Aybat:2011zv,Aybat:2011ge}].
It was demonstrated that TMD evolution [\refcite{Aybat:2011ta,Anselmino:2012aa}] appropriately takes into account the behavior of experimental data. One of the particularity of the TMD evolution consists in fact that
unlike usual collinear distributions where only collinear singularities are present, TMDs exhibit rapidity divergences 
along with collinear ones. Thus evolution is more intricate and describes not only how the form of distribution changes in 
terms of Bjorken-$x$ but also how the width is changed in momentum space $\bf k_T$. It was shown in Ref.~[\refcite{CollinsBook}] that TMD formalism in fact corresponds
to well known Collins-Soper-Sterman (CSS) resummation [\refcite{Collins:1981uw,Collins:1984kg}].

Evolution of twist-3 matrix elements was also recently worked out in Refs~[\refcite{Kang:2008ey,Zhou:2008mz,Vogelsang:2009pj,Braun:2009mi}] and the obtained result by three groups employing different methods
agree with each other [\refcite{Kang:2012em}]. The CSS resummation was also applied to spin dependent quantities in Ref.~[\refcite{Kang:2011mr}]. Along
with advances in TMD evolution implementation these results will lead to complete NLO knowledge of TMDs and twist-3 matrix elements which are sources
of spin asymmetries observed in different experiments in SIDIS, DY, and $e^+e^-$ annihilation.

\section*{Workshop}
This year we had 47 participants and 4 day scientific program consisted of 35 talks and discussions.

In particular evolution of twist-3 and TMD distributions was discussed in Refs.~[\refcite{Qiu,Kang,Cherednikov:2012ym,Gamberg}]. 
TMD factorization and evolution in the framework of Soft Collinear Effective Theory (SCET) was discussed
in Ref.~[\refcite{Echevarria:2012qe}]. Phenomenology of TMD evolution explored in Ref.~[\refcite{Melis}]. Results on TMDs from lattice QCD
were reported in Ref.~[\refcite{Engelhardt}]. Implementation of CSS formalism was reported in Ref.~[\refcite{Guzzi:2012jc}].

QCD in light front formalism was discussed in Ref.~[\refcite{Brodsky:2012je}] and connection of QCD vacuum structure to
intrinsic transverse momentum of partons was reported in Ref.~[\refcite{Weiss}]  
Effects of strong coupling running   in phenomenology was discussed in Ref.~[\refcite{Courtoy:2012hb}]. 

Several talks ~[\refcite{Liuti,Burkardt,Goldstein}] were dedicated to Generalized Parton Distributions and relation between TMDs and GPDs.
Singularities and basics of GPDs were reported in ~[\refcite{Radyushkin1,Radyushkin2}].

Relation of GPDs and TMDs to Wigner distribution and quark Orbital Angular Momentum was reported in Ref.~[\refcite{Lorce:2011tb}]. Generalized universality of TMDs
was discussed in Ref.~[\refcite{Buffing}].

Refs.~[\refcite{Metz,Pitonyak}] use multi-parton correlation amplitudes to describe Single Spin Asymmetries in Inclusive DIS and 
Double Spin Asymmetries in production of photons, hadrons and jets. Gluon TMDs were considered in Ref.~[\refcite{Mukherjee:2012mk}].

Next-to-Leading order $k_T$ factorization was reported in Ref.~[\refcite{Balitsky}]. Implementation of small-x formalism to DIS and proton-nuclei
scattering was discussed in Refs.~[\refcite{Chirilli:2012sk,Xiao}]. Color-Glass Condensate formalism was reported in  Ref.~[\refcite{Jalilian-Marian}].
New mechanism of creation SSA in High Energy QCD was reported in Ref.~[\refcite{Kovchegov:2012zx}].

Experimental results and future facilities were discussed in Refs.~[\refcite{Contalbrigo,Aschenauer,Turonski,Martin,Chen,Vossen}].

The workshop welcomes participation of students and we are glad that we had presentations [\refcite{Buffing,Pitonyak}] from students this year,
We will continue our commitment to encourage student's participation in the workshop in future. 

\section*{Acknowledgments}

We would like to acknowledge support of Jefferson Lab and help of Staff Services of Jefferson Lab without 
which the workshop could not be feasible. We would like to thank personally Marty Hightower,  
Cynthia Lockwood, Ruth Bizot, Stephanie Vermeire, and
MeLaina Evans for their help during the workshop.

Authored by a Jefferson Science Associate, LLC under U.S. DOE Contract 
No. DE-AC05-06OR23177. The U.S. Government retains a non-exclusive, 
paid-up, irrevocable, 
world-wide license to publish
or reproduce this manuscript for U.S. Government purposes.


\begin{thebibliography}{00}    

 \bibitem{proceedings}
{\em Int. J. Mod. Phys. Conf. Ser.} {\bf 4} (2011)  .

\bibitem{Dudek:2012vr}
J.~Dudek, R.~Ent, R.~Essig, K.~Kumar, C.~Meyer, {\it et al.},
\href{http://arxiv.org/abs/1208.1244}{{\tt arXiv:1208.1244 [hep-ex]}}.

\bibitem{Accardi:2011mz}
A.~Accardi, V.~Guzey, A.~Prokudin, and C.~Weiss,
\href{http://dx.doi.org/10.1140/epja/i2012-12092-7}{{\em Eur.Phys.J.} {\bf A48}
  (2012)  92}.

\bibitem{Abeyratne:2012ah}
S.~Abeyratne, A.~Accardi, S.~Ahmed, D.~Barber, J.~Bisognano, {\it et al.},
\href{http://arxiv.org/abs/1209.0757}{{\tt arXiv:1209.0757 [physics.acc-ph]}}.

\bibitem{Aubert:1985fx}
J.~Aubert {\it et al.}, {European Muon Collaboration} collaboration,
\href{http://dx.doi.org/10.1016/0550-3213(85)90635-2}{{\em Nucl.Phys.} {\bf
  B259} (1985)  189}.

\bibitem{Leader:1988vd}
E.~Leader and M.~Anselmino,
\href{http://dx.doi.org/10.1007/BF01566922}{{\em Z.Phys.} {\bf C41} (1988)
  239}.

\bibitem{deFlorian:2011ia}
D.~de~Florian, R.~Sassot, M.~Stratmann, and W.~Vogelsang,
\href{http://dx.doi.org/10.1016/j.ppnp.2011.12.027}{{\em Prog.Part.Nucl.Phys.}
  {\bf 67} (2012)  251--259}.

\bibitem{Efremov:1981sh}
A.~Efremov and O.~Teryaev,
{\em Sov.J.Nucl.Phys.} {\bf 36} (1982)  140.

\bibitem{Efremov:1984ip}
A.~Efremov and O.~Teryaev,
\href{http://dx.doi.org/10.1016/0370-2693(85)90999-2}{{\em Phys.Lett.} {\bf
  B150} (1985)  383}.

\bibitem{Qiu:1991pp}
J.-w. Qiu and G.~F. Sterman,
\href{http://dx.doi.org/10.1103/PhysRevLett.67.2264}{{\em Phys.Rev.Lett.} {\bf
  67} (1991)  2264--2267}.

\bibitem{Ji:2003ak}
X.-d. Ji,
\href{http://dx.doi.org/10.1103/PhysRevLett.91.062001}{{\em Phys.Rev.Lett.}
  {\bf 91} (2003)  062001}.

\bibitem{Wigner:1932eb}
E.~P. Wigner,
\href{http://dx.doi.org/10.1103/PhysRev.40.749}{{\em Phys.Rev.} {\bf 40} (1932)
   749--760}.

\bibitem{Kuraev:1977fs}
E.~Kuraev, L.~Lipatov, and V.~S. Fadin,
{\em Sov.Phys.JETP} {\bf 45} (1977)  199--204.

\bibitem{Balitsky:1978ic}
I.~Balitsky and L.~Lipatov,
{\em Sov.J.Nucl.Phys.} {\bf 28} (1978)  822--829.

\bibitem{Balitsky:1995ub}
I.~Balitsky,
\href{http://dx.doi.org/10.1016/0550-3213(95)00638-9}{{\em Nucl.Phys.} {\bf
  B463} (1996)  99--160}.

\bibitem{Kovchegov:1999yj}
Y.~V. Kovchegov,
\href{http://dx.doi.org/10.1103/PhysRevD.60.034008}{{\em Phys.Rev.} {\bf D60}
  (1999)  034008}.

\bibitem{Kovchegov:1999ua}
Y.~V. Kovchegov,
\href{http://dx.doi.org/10.1103/PhysRevD.61.074018}{{\em Phys.Rev.} {\bf D61}
  (2000)  074018}.

\bibitem{JalilianMarian:1997gr}
J.~Jalilian-Marian, A.~Kovner, A.~Leonidov, and H.~Weigert,
\href{http://dx.doi.org/10.1103/PhysRevD.59.014014}{{\em Phys.Rev.} {\bf D59}
  (1998)  014014}.

\bibitem{Iancu:2000hn}
E.~Iancu, A.~Leonidov, and L.~D. McLerran,
\href{http://dx.doi.org/10.1016/S0375-9474(01)00642-X}{{\em Nucl.Phys.} {\bf
  A692} (2001)  583--645}.

\bibitem{Ferreiro:2001qy}
E.~Ferreiro, E.~Iancu, A.~Leonidov, and L.~McLerran,
\href{http://dx.doi.org/10.1016/S0375-9474(01)01329-X}{{\em Nucl.Phys.} {\bf
  A703} (2002)  489--538}.

\bibitem{Schildknecht:2000zt}
D.~Schildknecht, B.~Surrow, and M.~Tentyukov,
\href{http://dx.doi.org/10.1016/S0370-2693(00)01397-6}{{\em Phys.Lett.} {\bf
  B499} (2001)  116--124}.

\bibitem{CollinsBook}
J.~C. Collins, {\em Foundations of Perturbative QCD}.
\newblock Cambridge University Press, 2011.

\bibitem{Aybat:2011zv}
S.~M. Aybat and T.~C. Rogers,
\href{http://arxiv.org/abs/1101.5057}{{\tt arXiv:1101.5057 [hep-ph]}}.

\bibitem{Aybat:2011ge}
S.~M. Aybat, J.~C. Collins, J.-W. Qiu, and T.~C. Rogers,
\href{http://dx.doi.org/10.1103/PhysRevD.85.034043}{{\em Phys.Rev.} {\bf D85}
  (2012)  034043}.

\bibitem{Aybat:2011ta}
S.~M. Aybat, A.~Prokudin, and T.~C. Rogers,
\href{http://dx.doi.org/10.1103/PhysRevLett.108.242003}{{\em Phys.Rev.Lett.}
  {\bf 108} (2012)  242003}.

\bibitem{Anselmino:2012aa}
M.~Anselmino, M.~Boglione, and S.~Melis,
\href{http://dx.doi.org/10.1103/PhysRevD.86.014028}{{\em Phys.Rev.} {\bf D86}
  (2012)  014028}.

\bibitem{Collins:1981uw}
J.~C. Collins and D.~E. Soper,
\href{http://dx.doi.org/10.1016/0550-3213(82)90021-9}{{\em Nucl.Phys.} {\bf
  B194} (1982)  445}.

\bibitem{Collins:1984kg}
J.~C. Collins, D.~E. Soper, and G.~F. Sterman,
\href{http://dx.doi.org/10.1016/0550-3213(85)90479-1}{{\em Nucl.Phys.} {\bf
  B250} (1985)  199}.

\bibitem{Kang:2008ey}
Z.-B. Kang and J.-W. Qiu,
\href{http://dx.doi.org/10.1103/PhysRevD.79.016003}{{\em Phys.Rev.} {\bf D79}
  (2009)  016003}.

\bibitem{Zhou:2008mz}
J.~Zhou, F.~Yuan, and Z.-T. Liang,
\href{http://dx.doi.org/10.1103/PhysRevD.79.114022}{{\em Phys.Rev.} {\bf D79}
  (2009)  114022}.

\bibitem{Vogelsang:2009pj}
W.~Vogelsang and F.~Yuan,
\href{http://dx.doi.org/10.1103/PhysRevD.79.094010}{{\em Phys.Rev.} {\bf D79}
  (2009)  094010}.

\bibitem{Braun:2009mi}
V.~Braun, A.~Manashov, and B.~Pirnay,
\href{http://dx.doi.org/10.1103/PhysRevD.80.114002}{{\em Phys.Rev.} {\bf D80}
  (2009)  114002}.

\bibitem{Kang:2012em}
Z.-B. Kang and J.-W. Qiu,
\href{http://dx.doi.org/10.1016/j.physletb.2012.06.021}{{\em Phys.Lett.} {\bf
  B713} (2012)  273--276}.

\bibitem{Kang:2011mr}
Z.-B. Kang, B.-W. Xiao, and F.~Yuan,
\href{http://dx.doi.org/10.1103/PhysRevLett.107.152002}{{\em Phys.Rev.Lett.}
  {\bf 107} (2011)  152002}.

\bibitem{Qiu}
J.-w. Qiu, {\em these proceedings}.

\bibitem{Kang}
Z.-B. Kang, {\em these proceedings}.

\bibitem{Cherednikov:2012ym}
I.~Cherednikov, T.~Mertens, and F.~Van~der Veken,
\href{http://arxiv.org/abs/1208.5410}{{\it these proceedings and }{\tt 
  arXiv:1208.5410 [hep-th]}}.

\bibitem{Gamberg}
L.~Gamberg, {\em these proceedings}.

\bibitem{Echevarria:2012qe}
M.~G. Echevarria, A.~Idilbi, and I.~Scimemi,
\href{http://arxiv.org/abs/1209.3892}{{\it these proceedings and }{\tt 
  arXiv:1209.3892 [hep-ph]}}.

\bibitem{Melis}
M.~Anselmino, E.~Boglione, and S.~Melis, {\em these proceedings}.

\bibitem{Engelhardt}
M.~Engelhardt, B.~Musch, P.~Haegler, J.~Negele, and A.~Schaefer, {\em these
  proceedings}.

\bibitem{Guzzi:2012jc}
M.~Guzzi and P.~M. Nadolsky,
\href{http://arxiv.org/abs/1209.1252}{{\it these proceedings and }{\tt 
  arXiv:1209.1252 [hep-ph]}}.

\bibitem{Brodsky:2012je}
S.~J. Brodsky and G.~de~Teramond,
\href{http://arxiv.org/abs/1208.3020}{{\it these proceedings and }{\tt 
  arXiv:1208.3020 [hep-ph]}}.

\bibitem{Weiss}
C.~Weiss, {\em talk at this workshop}.

\bibitem{Courtoy:2012hb}
A.~Courtoy and S.~Liuti,
\href{http://arxiv.org/abs/1208.5636}{{\it these proceedings and }{\tt 
  arXiv:1208.5636 [hep-ph]}}.

\bibitem{Liuti}
S.~Liuti, G.~Goldstein, J.~O.~G. Hernandez, and K.~Kathuria, {\em these
  proceedings}.

\bibitem{Burkardt}
M.~Burkrdt, {\em these proceedings}.

\bibitem{Goldstein}
G.~Goldstein, S.~Liuti, J.~O.~G. Hernandez, and K.~Kathuria, {\em these
  proceedings}.

\bibitem{Radyushkin1}
A.~V. Radyushkin, {\em these proceedings}.

\bibitem{Radyushkin2}
A.~V. Radyushkin, {\em these proceedings}.

\bibitem{Lorce:2011tb}
C.~Lorce and B.~Pasquini,
\href{http://arxiv.org/abs/1109.5864}{{\it these proceedings and }{\tt 
  arXiv:1109.5864 [hep-ph]}}.

\bibitem{Buffing}
M.~G.~A. Buffing and P.~J. Mulders, {\em these proceedings}.

\bibitem{Metz}
A.~Metz, D.~Pitonyak, A.~Schaefer, M.~Schlegel, W.~Vogelsang, {\it et al.},
  {\em these proceedings}.

\bibitem{Pitonyak}
D.~Pitonyak, {\em these proceedings}.

\bibitem{Mukherjee:2012mk}
A.~Mukherjee,
\href{http://arxiv.org/abs/1209.2774}{{\it these proceedings and }{\tt 
  arXiv:1209.2774 [hep-ph]}}.

\bibitem{Balitsky}
I.~Balitsky, {\em these proceedings}.

\bibitem{Chirilli:2012sk}
G.~A. Chirilli,
\href{http://arxiv.org/abs/1209.1614}{{\it these proceedings and }{\tt 
  arXiv:1209.1614 [hep-ph]}}.

\bibitem{Xiao}
G.~A. Chirilli, B.-W. Xiao, and F.~Yuan, {\em these proceedings}.

\bibitem{Jalilian-Marian}
J.~Jalilian-Marian, {\em these proceedings}.

\bibitem{Kovchegov:2012zx}
Y.~V. Kovchegov and M.~D. Sievert,
\href{http://arxiv.org/abs/1209.0727}{{\it these proceedings and }{\tt 
  arXiv:1209.0727 [hep-ph]}}.

\bibitem{Contalbrigo}
M.~Contalbrigo, {\em these proceedings}.

\bibitem{Aschenauer}
E.~Aschenauer, {\em talk at this workshop}.

\bibitem{Turonski}
P.~Nadel-Turonski, {\em talk at this workshop}.

 
\bibitem{Martin}
A.~Martin, {\em these proceedings}.

\bibitem{Chen}
J.-P. Chen, {\em these proceedings}.

\bibitem{Vossen}
A.~Vossen, {\em these proceedings}.

\end{thebibliography}

\newcommand{\href}[2]{#2}\begingroup\raggedright

\end{document}